\documentclass[preprint,12pt]{elsarticle}

\usepackage{amssymb}
\usepackage{amsmath}
\usepackage{colortbl}
\usepackage{multicol}
\usepackage{rotating}
\usepackage{color}
\usepackage{graphicx}

\journal{}

\begin{document}%\scriptsize
\begin{frontmatter}



\title{Link prediction based on path entropy}

\author[]{Zhongqi Xu}
\author[]{Cunlai Pu\corref{cor1}}
 \ead{pucunlai@njust.edu.cn}
 \cortext[cor1]{200 Xiaolingwei, Nanjing 210094, China. Tel: +8613915966537.}
 %\author[label1]{Siyuan Li}
\author[]{Jian Yang}
%\author{Zhongqi Xu, Cunlai Pu, Jian Yang}

\address{School of Computer Science and Engineering, Nanjing University of Science and Technology, Nanjing 210094, China}

\begin{abstract}
Information theory has been taken as a prospective tool for quantifying the complexity of complex networks. In this paper, we first study  the information entropy or uncertainty of a path using the information theory. Then we apply the path entropy to the link prediction problem in real-world networks. Specifically, we propose a new similarity index, namely Path Entropy (PE) index, which considers the information entropies of shortest paths between node pairs with penalization to long paths. Empirical experiments demonstrate that PE index outperforms the mainstream link predictors.
\end{abstract}


\end{frontmatter}


%\begin{multicols}{2}
\section{Introduction}
Fundamental principles underlying various complex systems, such as social, biological, technological systems, have attracted lots of attention from the network science community over the past two decades \cite{Dorog4,Barrat3,Newman1,Baraba2}. It has been demonstrated that plenty of real-world networks have scale-free degree distributions \cite{Baraba5,Baraba8,Cohen6,Albert7}, small-world effects \cite{Travers9,Watts10,Newman11,Hawick12}, and high clustering properties \cite{Klemm16,Bocca15}. Generally, social and collaboration networks are assortative mixing \cite{Newman17,Catan18}, while biological and technological networks are disassortative mixing \cite{Newman20,Zhou19}. Scale-free networks are very robust to random attacks, but are fragile to target attacks \cite{Albert21,Holme22,Pu23,Pu24}. Also, scale-free networks facilitate epidemic spreading since the epidemic threshold for scale-free networks approximates to zero \cite{Pastor25,Gang26,Pastor27,Yang28,Pu29}.
The deep understanding of network structures and dynamics helps us to make critical predictions of complex networks \cite{wenxu14}. Link prediction \cite{lv30,Wang31,Al34,Barzel32} is to estimate the existence possibility of links between unconnected nodes based on the network structures, node¡¯s attributes, and many others. Generally, there are two kinds of desired links: one is the missing links in the current network, and the other is the future links emerging in the evolution of the network. Link prediction has both scientific meaning and broad applications. On the one hand, prediction methods usually echo the fundamental organization rules of complex networks, and prediction performance in some sense indicates the predictability of complex networks \cite{Lv35}. For example, common neighborhood (CN)  based indices \cite{M37,G36,Liben38} are based on the high clustering property of complex networks. High prediction accuracy of CN-based indices indicates that the network has a strong clustering property and a large predictability. Preferential attachment (PA)   index \cite{Baraba5} reflects the rich-get-richer mechanism of social networks. In addition, link prediction provides us a natural standard for the comparison of various network models \cite{lv30}. On the other hand, link prediction is widely used in various applications, for example discovering potential interactions in protein-protein interaction networks \cite{LeiC39}, recommending goods and friends in social networks \cite{Al34,Sher41}, exploring coauthor relationships in collaboration networks \cite{NEWMAN01}, and even revealing hidden relations in terrorist networks \cite{Knoke43}.

Link prediction has long been discussed in computer science, but is booming recently in network science \cite{lv30}. The reason is that structural similarity indices are generally simpler with lower computational cost than machine learning based prediction methods. Specifically, structural similarity based methods can be classified into three groups: local indices \cite{M37,G36,Liben38,Baraba5,L45,T46}, global indices \cite{Katz48,Leicht49} and quasi-local indices \cite{L50,Liu51}. Local indices are usually defined by using the knowledge of common neighbors and node degree, which include CN, PA, Adamic-Adar (AA) \cite{L45}, resource allocation (RA) \cite{T46}, etc. Global indices are defined based on the whole network topological information, such as Katz Index \cite{Katz48}, Leicht-Holme-Newman (LHN) Index \cite{Leicht49}, and so on. Quasi-local indices are between local indices and global indices since the network topological information used in quasi-local indices is more than local indices, but less than global indices. Quasi-local indices contain local path (LP) index \cite{L50}, local random walk (LRW) index \cite{Liu51}, Superposed Random Walk (SRW) \cite{Liu51}, etc. Generally, the prediction accuracy of local indices is the lowest among the three groups of indices. However, the computational cost of local indices is the smallest among the three. Global indices are the opposite of local indices, while quasi-local indices fall in between. In addition, information of hierarchical  and community structures  \cite{Claus52,Soun53} has been referred to link prediction which further improves the prediction accuracy with additional computational cost.

Recently, information theory has been employed to quantify the complexity of complex networks structures with various scales \cite{Anand54,so55}. The Von Neumann entropy \cite{Passe56} and Shannon entropy \cite{Anand54} of a network are defined respectively. Bauer et al \cite{Baue57} used the maximum entropy principle in their construction of random graphs with arbitrary degree distribution. Bianconi \cite{Bianco58} studied the entropy of randomized network ensembles and found that network ensembles with fixed scale-free degree distribution have smaller entropy than that with homogeneous degree distribution. She \cite{Bianco59} further provided the expression of the entropy of multiplex networks ensembles. Halu et al \cite{Halu60} further studied the maximal entropy ensembles of spatial multiplex and spatial interacting networks. Entropy of network dynamics such as diffusion process \cite{Go61} and random walks \cite{Sina62} are also discussed. Network entropy measures have been applied to community detection \cite{Ros63}, aging and cancer progression characterization \cite{Meni64}, and very recently link prediction \cite{Tan65}.

So far, the information entropy or uncertainty embodied in a path has not yet been explored specifically. In complex networks, heterogeneity of paths can be further quantified by the path entropy or uncertainty. With path entropy, we can study how the path heterogeneity affects network properties and dynamics. In this paper, we firstly study the path entropy and obtain an approximate expression of path entropy which is based on the entropies of links in the path.
Then we apply  path entropy to the link prediction problems and propose a new similarity index based on path entropy. The outline of the article is as follows. Section 2 provides a detailed derivation of the entropy of a path. Section 3 gives the new similarity index. Section 4 presents the experiment results,  and section 5 provides the conclusion. There is also an appendix which introduces the basic link prediction framework and traditional similarity indices which are used in our experiments for comparison purpose.
\section{Information entropy of a path}
In information theory, the uncertainty of an event depends on the probability of its occurrence. Given an event $Q$ with occurrence possibility $P(Q)$, its information entropy or uncertainty $I(Q)$ is defined as \cite{Woodw66}:
\begin{equation}
I(Q)=-\log P(Q),
\end{equation}
where the base of the logarithm is 2, the same in the following. Apparently, the larger the occurrence possibility of event $Q$, the smaller the entropy of event $Q$ is.
For a node pair (a, b) in a network,  let's denote  $L_{ab}^1$ ($L_{ab}^0$), which means that there is (not) a link between a and b. Assuming there is no degree correlation among nodes in the network, the probability of  $L_{ab}^1$ is calculated as follows:
\begin{equation}
P(L^1_{ab})=1-P(L^0_{ab})=1-\prod_{i=1}^{k_b}{\frac{(M-{k_a})-i+1}{M-i+1}}=1-\frac{C_{M-{k_a}}^{k_b}}{C_M^{k_b}},
\end{equation}
where $k_a$ and $k_b$ are the degrees of $a$ and $b$. $M$ is the number of edges  in the network. Combing Eq. 1 and Eq. 2, we get the entropy of $L_{ab}^1$ as:
\begin{equation}
I(L^1_{ab})=-\log(P(L^1_{ab}))=-\log (1-\frac{C_{M-{k_a}}^{k_b}}{C_M^{k_b}}).
\end{equation}
Through the above derivation, we infer that $I(L^1_{ab})=I(L^1_{ba})$.  Assuming the network is sparse, we have $M\gg k_{max}$, where $k_{max}$ is the maximum node degree. Then, let's consider a simple path $D=v_0v_1{\cdots}v_\delta$ of length $\delta$. The occurrence probability of path $D$ is calculated as follows:
\begin{eqnarray}
P(D) \nonumber
    &=&P(L_{{v_0}{v_1}}^1,L_{{v_1}{v_2}}^1,\cdots,L_{{v_{\delta-1}}{v_\delta}}^1)\\\nonumber
    &=&P(L^1_{{v_0}{v_1}})P(L^1_{{v_1}{v_2}}|L^1_{{v_0}{v_1}})\cdots P(L^1_{{v_{\delta-1}}{v_\delta}}|(L^1_{{v_0}{v_1}},L^1_{{v_1}{v_2}},\cdots,L^1_{{v_{\delta-2}}{v_{\delta-1}}}))\\\nonumber
    &=&(1-P(L^0_{{v_0}{v_1}}))(1-P(L^0_{{v_1}{v_2}}|L^1_{{v_0}{v_1}})){\cdots}\\\nonumber
    &\quad&(1-P(L^0_{{v_{\delta-1}}{v_\delta}}|(L^1_{{v_0}{v_1}},L^1_{{v_1}{v_2}},\cdots,L^1_{{v_{\delta-2}}{v_{\delta-1}}})))\\\nonumber
    &=&(1-\frac{C_{M-k_{v_1}}^{k_{v_0}}}{C_M^{k_{v_0}}})(1-\frac{C_{M-1-{k_{v_2}}}^{{k_{v_1}}-1}}{C_{M-1}^{{k_{v_1}}-1}}){\cdots}(1-\frac{C_{M-(\delta-1)-{k_{v_\delta}}}^{k_{v_{\delta-1}}-1}}{C_{M-(\delta-1)}^{k_{v_{\delta-1}}-1}})\\\nonumber
    &=&(1-\frac{C_{M-k_{v_1}}^{k_{v_0}}}{C_M^{k_{v_0}}})\prod_{i=1}^{\delta-1}{(1-\frac{C_{M-i-{k_{v_{i+1}}}}^{{k_{v_i}}-1}}{C_{M-i}^{{k_{v_i}}-1}})}\\\nonumber
    &\approx&(1-\frac{C_{M-{k_{v_1}}}^{k_{v_0}}}{C_M^{k_{v_0}}})\prod_{i=1}^{\delta-1}{(1-\frac{C_{M-{k_{v_{i+1}}}}^{k_{v_i}}}{C_M^{k_{v_i}}})}\\\nonumber
    &\approx&\prod_{i=0}^{\delta-1}{(1-\frac{C_{M-{k_{v_{i+1}}}}^{k_{v_i}}}{C_M^{k_{v_i}}})}\\
    &\approx&\prod_{i=0}^{\delta-1}{P(L^1_{v_iv_{i+1}})}.
\end{eqnarray}

Eq. 4 means that the occurrence probability of a simple path approximates to the product of its links' occurrence probabilities. Then, the entropy of path $D$ is calculated as follows:
\begin{eqnarray}
I(D)\nonumber
&=&-\log(P(D))\\\nonumber
&\approx& -\log(\prod_{i=0}^{\delta-1}{P(L^1_{v_iv_{i+1}})})\nonumber  \\
&\approx& -\log(\prod_{i=0}^{\delta-1}(1-\frac{C_{M-{k_{v_{i+1}}}}^{k_{v_i}}}{C_M^{k_{v_i}}}))\nonumber \\
&\approx&\sum_{i=0}^{\delta-1}I(L^1_{v_iv_{i+1}}).
\end{eqnarray}
Eq. 5 indicates that the entropy of a path approximates to the sum of its links' entropies.

\section{Similarity index based on path entropy}
In link prediction, the link probability of an unconnected node pair is positively correlated with their topological similarities. Various similarity indices are proposed to quantify the topological similarity of node pairs \cite{lv30}. However, it is hard to find an universally  applied similarity index, since the essential organization rule or rules of various complex networks are usually unknown.

Information theory is a promising tool to measure the complexity of complex networks and has been applied in link prediction \cite{Tan65}. From the perspective of information theory, the link likelihood of a node pair is indicated by the link entropy. Large link entropy means that the node pair has a small probability to be connected with a link.  In link prediction problems, we are interested in the conditional entropy \cite{Tan65}:
\begin{equation}
I(L^1_{ab}|G')=I(L^1_{ab})-I(L^1_{ab};G'),
\end{equation}
where $G'$ is the part of the whole topological structure used in link prediction.  Generally, the conditional entropy decreases with the increase of amount of topological information used in the link prediction.
Here, we consider all the simple paths, and thus $G'=\bigcup\limits_{i=2}^l{\{D_{ab}^i}\}$, where $\{D^i_{ab}\}$ is the set of all simple paths of length $i$ between  $a$ and $b$. $l$ is the maximum length of simple paths we consider in the network.   Most of the path-based indices such as Katz and LP ignore the heterogeneity of paths. However, different paths  may make  different contributions to the link existence between the two end nodes.
Paths with large entropies are critical substructures for the network from the perspective of information theory, and the existence of these large-entropy paths greatly reduces the link entropy of the end nodes as indicated by Eq. 6. The contributions of all the simple paths are represented by $I(L^1_{ab};G')$ in Eq. 6, which is approximately calculated as follows:
\begin{equation}
I(L^1_{ab};G')=I(L^1_{ab};\bigcup_{i=2}^l{\{D_{ab}^i\}}){\approx}\sum_{i=2}^l{\{\frac{1}{i-1}(\sum_{D\in\{D_{ab}^i\}}I(D))\}},
\end{equation}
where $1/(i-1)$ is the weight of simple paths with length $i$, with which the contributions of long paths are penalized. This is based on the common assumption that  the longer the path is, the less important the path is in link prediction. Note that we have also checked other weight forms such as $1/i$, $1/\sqrt{(i)}$, etc., and find that $1/(i-1)$ corresponds to the best prediction performance.
For two unconnected nodes, large link entropy  means small link probability, or in some sense small similarity.
Our path-entropy (PE) based similarity index is defined as the negative of the conditional entropy as follows:
 \begin{eqnarray}
S_{ab}^{PE}&=&-I(L^1_{ab}|\bigcup_{i=2}^l{\{D_{ab}^i\}})\nonumber \\
           &=&I(L_{ab}^1;\bigcup_{i=2}^l{\{D_{ab}^i}\})-I(L_{ab}^1) \nonumber \\
           &=&\sum_{i=2}^l{\{\frac{1}{i-1}(\sum_{D\in\{D_{ab}^i\}}I(D))\}}-I(L_{ab}^1) \nonumber \\
           &=&\sum_{i=2}^l{\{\frac{1}{i-1}(\sum_{D\in\{D_{ab}^i\}}\{\sum_{j=0}^{i-1}I(L^1_{v_jv_{j+1}})\})\}}-I(L_{ab}^1) \\
           &=&\sum_{i=2}^l{\{\frac{1}{i-1}(\sum_{D\in\{D_{ab}^i\}}\{\sum_{j=0}^{i-1}{log(\frac{C_M^{k_{v_j}}}{C_M^{k_{v_j}}-C_{M-{k_{v_{j+1}}}}^{k_{v_j}}})\})}\}}
           -log(\frac{C_M^{k_a}}{C_M^{k_a}-C_{M-{k_b}}^{k_a}})  \nonumber
\end{eqnarray}

\begin{figure}
 \centering
%\onefigure{F2.eps}
\includegraphics[width=1.6in,height=1.5in]{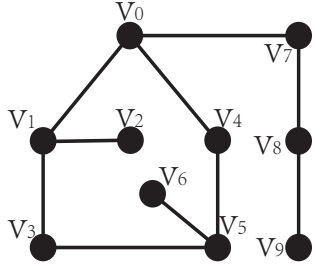}
\caption{A simple network for illustration purpose. Assuming $l=3$, we obtain $S_{v0v5}^{PE}>S_{v0v3}^{PE}>S_{v0v8}^{PE}>S_{v0v9}^{PE}>S_{v0v2}^{PE}>S_{v0v6}^{PE}$ based on PE index.}
%\label{fig.2}
\end{figure}

We illustrate the calculation of PE with a simple network  as shown in Figure 1. We set $l$ as 3. Considering node pair $(v0,v3)$, we have $\{D_{v0v3}^2\}=\{v0v1v3\}$ and $\{D_{v0v3}^3\}=\{v0v4v5v3\}$. Based on Eq. 3, we have $I(L_{v0v1}^1)=0.497$, and $I(L_{v0v3}^1)=I(L_{v0v4}^1)=I(L_{v4v5}^1)=I(L_{v1v3}^1)=I(L_{v3v5}^1)=0.907$. Then, we get $I(D_{v0v3}^2)=1.404$ and $I(D_{v0v3}^3)=2.721$ using Eq. 5. Finally, we obtain $S_{v0v3}^{PE}=1.858$ based on Eq. 8. Similarly, we get $S_{v0v2}^{PE}=0.497$, $S_{v0v5}^{PE}=2.473$, $S_{v0v6}^{PE}=0.039$, $S_{v0v8}^{PE}=1.404$, and $S_{v0v9}^{PE}=0.580$. Thus, we have $S_{v0v5}^{PE}>S_{v0v3}^{PE}>S_{v0v8}^{PE}>S_{v0v9}^{PE}>S_{v0v2}^{PE}>S_{v0v6}^{PE}$.
Note that node pair $(v0,v9)$ have no common neighbors, but based on PE index their possibility of being connected is larger than node pair $(v0,v2)$ which have one common neighbor. This indicates that the contribution of path $v0v7v8v9$ is greater than that of path $v0v1v2$, given that  node pairs $(v0,v9)$ and $(v0,v2)$ have the same link entropy($I(L_{v0v9}^1)=I(L_{v0v2}^1)$).

\section{Results }
\begin{table}
\resizebox{\textwidth}{!}{ %
\centering
\begin{tabular}{lccccccc}
\hline
Networks&$|V|$&$|E|$&$<d>$&$<k>$&$H$&$C$&$r$ \\ \hline
\rowcolor[gray]{.7}
C.elegans&297&2148&2.46&14.4646&1.8008&0.3079&-0.163\\
\rowcolor[gray]{.9}
FWFW&128&2075&1.78&32.4219&1.2370&0.3346&-0.112 \\
\rowcolor[gray]{.7}
SmaGri&1024&4916&2.98&9.6016&3.9475&0.3486&-0.193 \\
\rowcolor[gray]{.9}
Poliblogs&1222&16714&2.74&27.3552&2.9707&0.3600&-0.221 \\
\rowcolor[gray]{.7}
Power&4941&6594&18.99&2.6691&1.4504&0.1065&0.004 \\
\rowcolor[gray]{.9}
Router&5022&6258&6.45&2.4922&5.5031&0.0329&-0.138 \\

\rowcolor[gray]{.7}
Yeast&2375&11693&5.09&9.8467&3.4756&0.3883&0.454 \\
\rowcolor[gray]{.9}
Email&1133&5451&3.60&9.6222&1.9421&0.2540&0.078 \\
\rowcolor[gray]{.7}
Kohonen&3704&12673&3.67&6.8429&9.3170&0.3044&-0.121 \\
\rowcolor[gray]{.9}
EPA&4253&8897&4.50&4.1839&6.7668&0.1360&-0.304 \\
\rowcolor[gray]{.7}
SciMet&2678&10368&4.18&7.7431&2.4265&0.2026&-0.035 \\
\hline \\
\end{tabular}}
\caption{The  topological statistics of eleven real networks. $|v|$ is the number of nodes. $|E|$ represents the number of links.  $<d>$ is the average distance. $<k>$ is the average degree.  $H$ represents the degree heterogeneity, defined as $H=\frac{<k^2>}{{<k>}^2}$. $C$ and $r$ are the clustering coefficient and assortativity coefficient respectively.}
\end{table}

We compare our similarity index with  six mainstream indices (see Appendix A) on eleven real-world networks \cite{Watts10,FWFW68,SmaGri69,Poliblogs70,Router71,Yeast72,email73}, the statistics of which are summarized in Table 1. In those networks, directed links  are taken as undirected links by ignoring the directions, and self-connections are removed. For unconnected networks, we choose their largest connected components.  The data of those networks comes from disparate fields. $\romannumeral1)$ C.elegans \cite{Watts10}:  The neural
network of the nematode Caenorhabditis elegans. $\romannumeral2)$ FWFW \cite{FWFW68}: The food web of Florida ecosystem. $\romannumeral3)$ SmaGri \cite{SmaGri69}: The network composed of citations to Small $\&$ Griffith and Descendants. $\romannumeral4)$ Poliblogs \cite{Poliblogs70}: A network of the US political blogs. $\romannumeral5)$ Power \cite{SmaGri69}: The electrical power grid of the western US. $\romannumeral6)$ Router \cite{Router71}: The router-level topology of the internet. $\romannumeral7)$ Yeast \cite{Yeast72}: The protein-protein interaction network of yeast. $\romannumeral8)$ Email \cite{email73}: The network of e-mail interchanges between members of the Univeristy Rovira i Virgili (Tarragona).
$\romannumeral9)$ Kohonen \cite{SmaGri69}: A network of articles with topic self-organizing maps or references to Kohonen T. $\romannumeral10)$ EPA \cite{SmaGri69}: A network of web pages linking to the website www.epa.gov. $\romannumeral11)$ SciMet \cite{SmaGri69}: A network of articles from or citing Scientometrics.

Introductions of AUC and Precision are shown in Appendix A. The division of the training and test sets is also provided in Appendix A. The results of AUC and Precision for various similarity indices are provided  in Table 2 and  3 respectively.  For our PE index,  $l=2$ and $l=3$ are considered in the experiment.  Note that all the results are the  average over $100$ independent runs.

\begin{table}
\resizebox{\textwidth}{!}{ %
\begin{tabular}{lcccccccc}
\hline
\textbf{Nets$\diagdown$Index}&CN&RA&AA&LNB-CN&LNB-RA&MI&PE($l=2$)&PE($l=3$) \\ \hline
\rowcolor[gray]{.7}
C.elegans&0.8503&0.8712&0.8670&0.8620&0.8671&0.8411&0.8715&0.8849\\
\rowcolor[gray]{.9}
FWFW&0.6071&0.6129&0.6090&0.6215&0.6212&0.5023&0.5684&0.8617 \\
\rowcolor[gray]{.7}
SmaGri&0.8485&0.8586&0.8587&0.8582&0.8587&0.9030&0.9028&0.9172 \\
\rowcolor[gray]{.9}
Poliblogs&0.9241&0.9286&0.9275&0.9264&0.9283&0.9324&0.9336&0.9458 \\
\rowcolor[gray]{.7}
Power&0.6265&0.6265&0.6265&0.6262&0.6264&0.6081&0.6430&0.6904 \\
\rowcolor[gray]{.9}
Router&0.6534&0.6537&0.6536&0.6535&0.6533&0.9570&0.9590&0.9866 \\

\rowcolor[gray]{.7}
Yeast&0.9166&0.9176&0.9173&0.9166&0.9174&0.9377&0.9456&0.9750 \\
\rowcolor[gray]{.9}
Email&0.8556&0.8572&0.8576&0.8566&0.8567&0.8905&0.9035&0.9211 \\
\rowcolor[gray]{.7}
Kohonen&0.8278&0.8355&0.8356&0.8354&0.8356&0.9109&0.9145&0.9312 \\
\rowcolor[gray]{.9}
EPA&0.6099&0.6111&0.6112&0.6125&0.6124&0.9254&0.9209&0.9496 \\
\rowcolor[gray]{.7}
SciMet&0.7983&0.7997&0.8001&0.7998&0.7995&0.8715&0.8824&0.9216 \\
\hline \\
\end{tabular}}
\caption{Prediction accuracy measured by AUC on eleven real-world networks.}
\end{table}
Table 2 shows that for AUC, PE with $l=2$ already achieves better performance than the other mainstream similarity indices except  FWFW.  When $l=3$, PE gets better performance than $l=2$, and  AUC for FWFW is greatly improved.  This is generally because for PE index the more topological information used in link prediction, the better the prediction performance is. However, considering the contributions of long simple paths, which is relatively small, and the large computational cost they cause, it is reasonable to just consider short simple paths in link prediction.
 Note that for Poliblogs and Yeast, AUC reach more than $90\%$, so it is difficult to improve their prediction accuracy. Also, the small average node degree and large average distance  for Power limit the prediction accuracy.

 Table 3 shows that for precision $l=2$ for PE is not enough, since the corresponding precision values are not better than the mainstream similarity indices. When $l=3$, the precision values of PE are generally larger than the other indices. The exceptions are Power and Email(bold in Table 3) for which the Precision values of  PE with $l=3$ are even worse than PE with $l=2$.

\begin{table}
\resizebox{\textwidth}{!}{ %
\begin{tabular}{lcccccccc}
\hline
\textbf{Nets$\diagdown$Index}&CN&RA&AA&LNB-CN&LNB-RA&MI&PE($l=2$)&PE($l=3$) \\ \hline
\rowcolor[gray]{.7}
C.elegans&0.1192&0.1295&0.1311&0.1315&0.1291&0.1421&0.1262&0.1841\\
\rowcolor[gray]{.9}
FWFW&0.0856&0.0890&0.0913&0.1058&0.1084&0.0570&0.0242&0.4729 \\
\rowcolor[gray]{.7}
SmaGri&0.1741&0.1909&0.2015&0.2056&0.1965&0.2381&0.2185&0.2305 \\
\rowcolor[gray]{.9}
Poliblogs&0.4346&0.2470&0.3729&0.4114&0.2601&0.4785&0.3377&0.5365 \\
\rowcolor[gray]{.7}
Power&0.1068&0.0810&0.1027&0.1637&0.0947&0.1741&0.1271&\textbf{0.1186} \\
\rowcolor[gray]{.9}
Router&0.1031&0.0867&0.1223&0.1182&0.0654&0.2168&0.1099&0.6049 \\

\rowcolor[gray]{.7}
Yeast&0.6755&0.4960&0.7042&0.6951&0.6207&0.8159&0.6862&0.8624 \\
\rowcolor[gray]{.9}
Email&0.3046&0.2561&0.3150&0.3215&0.2644&0.3260&0.3461&\textbf{0.2234} \\
\rowcolor[gray]{.7}
Kohonen&0.1448&0.1318&0.1438&0.1634&0.1427&0.2379&0.1972&0.2432 \\
\rowcolor[gray]{.9}
EPA&0.0154&0.0401&0.0358&0.0253&0.0414&0.0565&0.0198&0.4617 \\
\rowcolor[gray]{.7}
SciMet&0.1535&0.1224&0.1347&0.1502&0.1221&0.1629&0.1487&0.2344 \\
\hline \\
\end{tabular}}
\caption{Prediction accuracy measured by Precision (top-$100$) on eleven real-world networks.}
\end{table}

\section{Conclusion}
In summary, we quantitatively study the influence of paths in link prediction by using the information theory. We obtain that the information entropy of a path is approximately equal to the sum of its links¡¯ information entropies. Path entropy is a natural metric for quantifying the structure importance of a path in the network. We apply path entropy in link prediction problems, and propose a new similarity index.  Our similarity index considers the contributions of all simple paths in link prediction measured by path entropies with penalty to long paths. Simulation results on real-world networks demonstrate that our index generally outperforms the other mainstream similarity indices with higher prediction accuracy measured by AUC and Precision.
The reason is that most of the other similarity indices consider the number of common neighbors, node degrees, path lengths, etc. However, these metrics are relatively coarse compared to those metrics in the information theory framework. With path entropy, we better quantify the role of paths in link prediction, and thus can design more efficient link predictors. We also believe that path entropy can be applied to other network problems such as epidemic spreading, network attacks and so on.

\begin{appendix}
\section{Problem description and standard metrics}
  Assuming an undirected and unweighted network $G(V, E)$, where $V$ and $E$ are the sets of nodes and links respectively.  Clearly, $G$ has $|V|(|V|-1)/2$ node pairs totally, which constitute the universal set $U$. To measure the performance of similarity indices in link prediction,  $E$ is randomly divided into two parts: a training set $E^T$ and a test set $E^P$. In our experiment, $E^T$ and $E^P$ are generated with the 90/10  rule \cite{lv30}.  Obviously, $E^T \cup E^P=E$ and $E^T \cap E^P=\emptyset$.

Two standard metrics AUC and Precision are often used in link prediction. AUC is the area under the receiver operating characteristic (ROC)  curve.
  When calculating AUC, each node pair in $U-E^T$ is given a similarity score based on a given  similarity index. Then, each time we randomly pick a  link from $E^P$ and a nonexistence link from $U-E$ and compare their scores. If among $n$ times of independent comparisons, there are $n'$ times that the score of the  link from $E^P$  is higher than the link from $U-E$,  and $n''$ times that they have the same scores, then AUC is calculated as:
 \begin{equation}
AUC=\frac{n'+0.5n''}{n}.
\end{equation}
Apparently, AUC should be close to $0.5$ if the scores are assigned from an independent and identical distribution. Therefore, an AUC larger than 0.5 means the link prediction method is better than pure chance, and similarity index with the larger AUC is always preferable.
Precision cares about the prediction accuracy of   top ranked  links.  If among top $L$ links ranked by similarity scores, there are $m$ links belonging to $E^P$, then Precision is calculated as:
 \begin{equation}
Precision=\frac{m}{L}.
\end{equation}

We here introduce the  six mainstream similarity indices which are used for comparison purpose in our experiment.

(1)Common neighbors (CN) \cite{M37}. This index defines the similarity score of two nodes as  the number of their common neighbors, which is:
 \begin{equation}
S_{ab}=|\Gamma(a)\cap\Gamma(b)|=|O_{ab}|,
\end{equation}
where $\Gamma(a)$ is the set of neighbors of $a$, and $O_{ab}$ is the set of common neighbors of $a$ and $b$.

(2)Resource Allocation (RA) \cite{T46}. This index  considers the degree of common neighbors with penalty to large degree nodes, which is:
 \begin{equation}
S_{ab}=\sum_{z\in{O_{ab}}}{\frac{1}{|\Gamma(z)|}}.
\end{equation}

(3)Adamic-Adar Index (AA) \cite{L45}. This index is similar to RA, but considers the logarithm of node degree, which is:
\begin{equation}
S_{ab}=\sum_{z\in{O_{ab}}}{\frac{1}{log(|\Gamma(z)|)}}.
\end{equation}

(4)Local Na\"{i}ve Bayes form of CN (LNB-CN) \cite{Liu74}. This index  weights the contributions of common neighbors by using the  Na\"{i}ve Bayes model, which is defined as follows:
\begin{equation}
S_{ab}=|O_{ab}|log\eta+\sum_{z\in{O_{ab}}}{logR_z},
\end{equation}
where $\eta=\frac{|V|(|V|-1)}{2|E^T|}-1$,  and $R_z=\frac{N_{{\Delta}z}+1}{N_{{\Lambda}z}+1}$.   $N_{{\Delta}z}$ and $N_{{\Lambda}z}$ are respectively the numbers of connected and disconnected node pairs whose common neighbors include $z$.

(5)Local Na\"{i}ve Bayes form of RA (LNB-RA) \cite{Liu74}. Similar to LNB-CN, this index combines the RA index with the  Na\"{i}ve Bayes model, defined as:
\begin{equation}
S_{ab}=\sum_{z\in{O_{ab}}}{\frac{1}{|\Gamma(z)|}(log\eta+logR_z)}.
\end{equation}

(6)Mutual Information index (MI) \cite{Tan65}. This index quantifies the contributions of  common neighbors with the mutual information theory, defined as:
\begin{equation}
S_{ab}=\sum_{z\in{O_{ab}}}{I(L_{ab}^1;z)}-I(L_{ab}^1),
\end{equation}
where $I(L_{ab}^1)$ is calculated with Eq. 3. $I(L_{ab}^1;z)$ is calculated as follows:
\begin{equation}
I(L_{ab}^1;z)\approx \frac{1}{|\Gamma(z)|(|\Gamma(z)-1|)}\sum_{m, n\in \Gamma(z), m\neq n}(I(L_{mn}^1)+\log \frac{N_{\Delta z}}{N_{\Delta z}+N_{\Lambda z}}).
\end{equation}
\end{appendix}

%\label{}





%\end{multicols}
\end{document}